\documentclass[aps,prl,superscriptaddress,twocolumn,showpacs,amsmath,amssymb,letter]{revtex4}

\usepackage{graphicx}
\usepackage{dcolumn}
\usepackage{bm}

\begin{document}

\title{Anomalous Fermi-Surface Dependent Pairing in a Self-Doped High-T$_{c}$ Superconductor}

\author{Yulin Chen}
\affiliation {Department of Physics, Applied Physics, and Stanford
Synchrotron Radiation Laboratory, Stanford University, Stanford,
CA 94305, USA}

\author{Akira Iyo}
\affiliation {National Institute of Advanced Industrial Science
and Technology, Tsukuba, Ibaraki, 305-8568, Japan}

\author{Wanli Yang}
\affiliation {Department of Physics, Applied Physics, and Stanford
Synchrotron Radiation Laboratory, Stanford University, Stanford,
CA 94305, USA}

\author{Xingjiang Zhou}
\affiliation {Department of Physics, Applied Physics, and Stanford
Synchrotron Radiation Laboratory, Stanford University, Stanford,
CA 94305, USA}

\author{Donghui Lu}
\affiliation {Department of Physics, Applied Physics, and Stanford
Synchrotron Radiation Laboratory, Stanford University, Stanford,
CA 94305, USA}

\author{Hiroshi Eisaki}
\affiliation {National Institute of Advanced Industrial Science
and Technology, Tsukuba, Ibaraki, 305-8568, Japan}

\author{Thomas P. Devereaux}
\affiliation {Department of Physics, University of Waterloo,
Waterloo, ON, Canada N2L 3G1}

\author{Zahid Hussain}
\affiliation {Advanced Light Source, Lawrence Berkeley National
Laboratory, Berkeley California 94720, USA}

\author{Z.-X. Shen}
\affiliation {Department of Physics, Applied Physics, and Stanford
Synchrotron Radiation Laboratory, Stanford University, Stanford,
CA 94305, USA}

\date{\today}

\begin{abstract}
We report the discovery of a self-doped multi-layer high T$_{c}$
superconductor Ba$_{2}$Ca$_{3}$Cu$_{4}$O$_{8}$F$_{2}$ (F0234)
which contains distinctly different superconducting gap magnitudes
along its two Fermi surface(FS) sheets. While formal valence
counting would imply this material to be an undoped insulator, it
is a self-doped superconductor with a T$_{c}$ of 60K, possessing
simultaneously both electron- and hole-doped FS sheets.
Intriguingly, the FS sheet characterized by the much larger gap is
the electron-doped one, which has a shape disfavoring two
electronic features considered to be important for the pairing
mechanism: the van Hove singularity and the antiferromagnetic
$(\pi/a,\pi/a)$ scattering.
\end{abstract}


\maketitle

The origin of the very high superconducting transition temperature
(T$_{c}$) in the ceramic copper oxide superconductors is often
quoted as one of the great mysteries in modern physics. Important
insights on high-T$_{c}$ superconductivity are often gained
through investigations on new compounds with unusual properties.
The recently synthesized single crystalline
Ba$_{2}$Ca$_{3}$Cu$_{4}$O$_{8}$F$_{2}$(F0234) is one such example.
While valence charge counting based on the canonical chemical
formula puts Cu valence as 2$^+$ thus the material as a
half-filled Mott insulator, the compound turns out to be a
superconductor with T$_{c}$ of 60K\cite{samplegrow1,samplegrow2}.
We performed high resolution angle-resolved photoemission
spectroscopy(ARPES) measurements on this compound. Our data reveal
at least two metallic Fermi surface(FS) sheets with volumes
equally above and below half-filling. We also found that the two
FS pieces split significantly along the nodal direction which is
very different from other multilayer cuprates\cite{Andrea_review}.
Most interestingly, we found an anomalous FS dependence of the
superconducting gap in which the larger gap associates with the
most bonding FS that lies further away from the antiferromagnetic
reciprocal lattice zone boundary and the van Hove points. This
discovery puts a strong constraint on theory.

\begin{figure}
\includegraphics[width=0.35\textwidth]{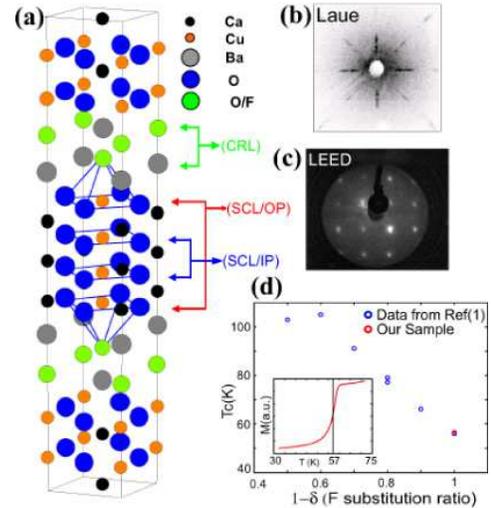}
\caption{\label{fig:epsart} (a): Crystal structure of
Ba$_{2}$Ca$_{3}$Cu$_{4}$O$_{8}$(O$_\delta$F$_{1-\delta}$)$_2$.
There are four CuO$_{2}$ layers in a unit cell with the outer two
having apical F atoms (b): Laue pattern of the sample shows the
tetragonal symmetry without superstructure. (c): LEED pattern on
the cleaved surface after ARPES measurement confirms no surface
reconstruction. (d): T$_{c}$ v.s. F-substitution ratio with SQUID
measurement of the measured sample(inset).}
\end{figure}

The crystal structure of
Ba$_{2}$Ca$_{3}$Cu$_{4}$O$_{8}$(O$_\delta$F$_{1-\delta}$)$_2$ is
shown in Fig. 1a. It has tetragonal symmetry with alternate
stacking of superconducting layers (SCLs) and charge reservoir
layers (CRLs). Within a conventional unit cell, there are four
CuO$_{2}$ SCLs that can be divided into two crystallographically
inequivalent groups: the outer pair of CuO$_{2}$ planes (OP) with
apical F atoms and the inner pair (IP) without. As shown in Fig.
1d, depending on the apical O/F substitution ratio, the
superconducting transition temperature (T$_{c}$) varies and can
reach 105K when $\delta$=0.4\cite{samplegrow2}.
F0234($\delta\approx$ 0), the material studied in this
investigation, has a T$_{c}\approx$ 60K.

ARPES experiments were carried out on F0234 samples by using 55eV
photons as described before\cite{spacecharge}. The single
crystalline samples were grown under high
pressure\cite{samplegrow2}.  Total (convolved) energy resolution
of the measurements was 16meV; and the angular resolution was 0.2
degrees. Spectra presented in this paper were measured at 20K and
80K, and the Fermi energy (E$_{f}$) is internally referenced to
the leading edge position at the d-wave node within each set of
data.

\begin{figure}
\includegraphics[width=0.41\textwidth]{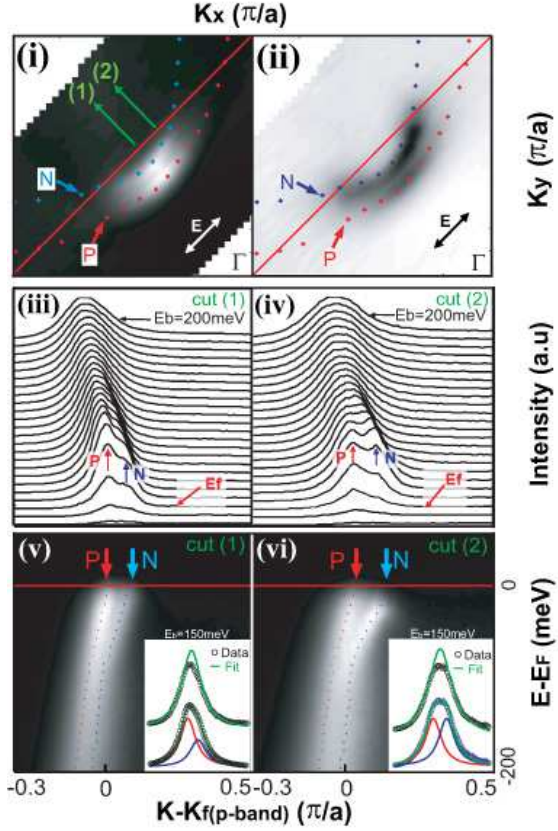}
\caption{\label{fig:epsart} (i): FS map(T=20K) by integrating
spectral intensity from E$_{b}$=20meV to E$_{f}$. White
double-arrow shows photon polarization. Red and blue symbols give
FS contours of band P and N extracted by method described in text.
(ii): Derivative map by integrating the derivative of the spectral
density from E$_{b}$=20meV to E$_{f}$. Red and blue symbols are
defined as in (i). (iii$\sim$vi): MDC(iii, iv) and image(v, vi)
plots of raw data along cuts marked by green arrows in (i). Arrows
in (iii,iv) show the P and N contributions in MDC curves. Band
maxima (k$_{f}$) positions are indicated by arrows in (v, vi)
marked as P and N respectively. Dot lines superimposed on (v, vi)
are MDC fit dispersions of both bands. Insets in (v, vi)
illustrate one(above) and two(below) Lorentzian peaks fits of the
MDCs at E$_{b}$=150meV.}
\end{figure}

Fig. 2 shows data recorded at 20K that revealed two sets of
distinct bands and FS sheets.  The spectral intensity map in Fig.
2(i) is a typical way to illustrate the FS topology. Due to the
opening of the superconducting gap, the energy integration window
is chosen from 20meV binding energy (E$_{b}$) to Fermi energy
(E$_{f}$). At the first glance at plot (i), it looks similar to
other cuprate
superconductors\cite{Marshall,Bi_HDing,Borisenko_FS,Yoshida,Andrea_review},
which in general show hole like topology around $(¡À\pi/a,
¡À\pi/a)$ in k-space, and the intensity peaks along
$(0,0)-(\pi/a,\pi/a)$ nodal direction. However, a closer look at
the data reveals the presence of at least two dispersive bands
(marked as P and N respectively), which are clearly discernable in
both the momentum distribution curves (MDCs) stack plots (iii, iv)
and the image plots (v, vi) from data cuts illustrated by green
arrows in plot (i). From (iii)$\sim$(vi), we see that the double
peak structures bifurcate further when the two bands disperse to
lower binding energy until being limited by the superconducting
gap which is more pronounced for band N. The MDC derived
dispersions, obtained by fitting the MDC with two Lorentzian peaks
plus linear background, are superimposed on plot (v, vi),
revealing dispersion kinks at E$_{b}\sim$85meV. We see that the
kink of band N is stronger than that of band P; and both are
stronger in cut 2 (plot vi) than in cut 1 (plot v). To show the
validity of the fitting procedure at high binding energy, in inset
of Fig.2(v,vi), we illustrate one Lorentzian (above) and two
Lorentzian peaks(below) fits for MDC at E$_{b}$=150meV. It's clear
that one Lorentzian peak can't fit the MDC line shape well.

The presence of the two bands can be seen, but not very distinct
in Fig. 2(i) due to the large gap magnitude of band N in the
region away from the node and the overlapping with the P band
close to the node. As another mapping technique, Fig. 2(ii) shows
a derivative map by integrating the derivative of the spectra
intensity within the same energy window as in (i). Because the
derivative emphasizes the spectral weight change at the leading
edge, it is more sensitive in detecting band-top positions, thus
the FS contours. As expected, the two bands' FS contours become
more distinct in Fig. 2(ii).

\begin{figure}
\flushleft
\includegraphics[width=0.45\textwidth]{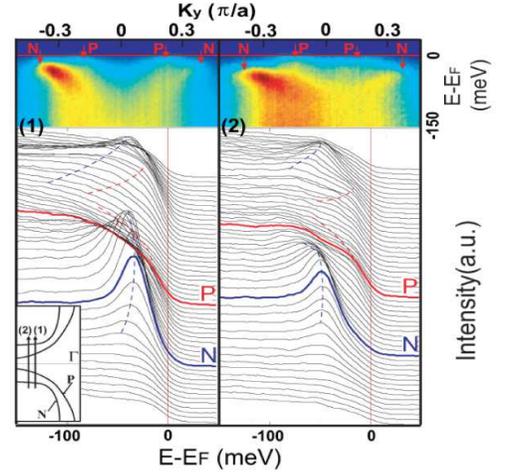}
\caption{\label{fig:epsart} Raw data image (top) and EDC stack
(bottom) plots along cuts indicated in the inset. Red and blue
broken lines in EDC plots are guides to the eye that follow the
band dispersions. Red and blue bold EDCs represent the EDCs at
k$_{f}$ positions for a pair of P and N bands, which are also
indicated by arrows in the image plots.}
\end{figure}

The same behavior can be seen in Fig. 3 with data taken at another
measuring geometry (see inset). The presence of two sets of bands
is self-evident in the raw data image plot and the energy
distribution curve (EDC) stack plot in both k-space quadrants. We
highlight the EDCs at the band maxima (k$_{f}$) positions for one
set of bands in the EDC stack plots for clarity. It is obvious
from these data that the gap of band N is larger than that of band
P as we will discuss in detail later.

To quantitatively determine the FS contours of both bands, we use
EDC and MDC analysis complementarily\cite{MDC}. We first fit every
EDC in each cut to locate the k$_{f}$ positions for both bands
based on the minimum superconducting gap criteria\cite{FS_HDing}
then cross check by fitting MDCs for consistency.  The FS contours
acquired from Sample 1 are superimposed on Fig.2(i, ii), and those
from sample 2 are shown in Fig.5a. If we calculate the FS volumes
enclosed by each FS sheet with respect to $(\pi/a,\pi/a)$ point
within each quadrant(see Fig. 5a), we get 0.61$\pm$0.04 (unit
$(\pi/a)^2$) for band P FS and 0.4$\pm$0.03 for band N FS.

In the conventional band structure framework, both FS sheets are
hole-like. The presence of multiple FS sheets is caused by
hybridization of bands from different CuO$_{2}$ planes. Similar to
an earlier investigation on another multilayer
system\cite{Donglai}, the number of the bands observed is less
than expected from the number of CuO$_{2}$ planes. The fact that
we only see two FS sheets rather than four as expected indicates
that the additional splitting may be too small to be resolved.
Compared to what was observed in Bi$_{2}$Sr$_{2}$CaCu$_{2}$O$_{8}$
(Bi2212)\cite{Donglai2, Chuang}, we find that the splitting of the
FS around the nodal direction is significantly larger.

However, in terms of doped Mott insulator description as is
generally used in the field, if we compare our result to the
half-filling state in which the FS should have an area of 0.5 in
each quadrant, we find that the band P is "hole-doped" and the
band N is "electron-doped" -- the reason for them being marked as
P and N respectively. Under the convention established for the
cuprates, these two pieces of FS are approximately 20$\pm$8\%
hole- and 20$\pm$6\% electron-doped respectively although the
nominal composition indicates F0234$(\delta=0)$ to be an undoped
Mott insulator. This is the first self-doping case observed in the
cuprates.

The self-doping behavior, or significant FS splitting due to
inter-CuO$_{2}$ plane interaction, while apparently surprising, is
consistent with the band structure calculations where the areas of
various FS' in multilayer materials are often found to be very
different\cite{Pickett,Andersen,Hamada}. Layer dependent doping
has been observed in other multilayer high T$_{c}$ superconductors
such as HgBa$_{2}$Ca$_{n}$Cu$_{n+1}$O$_{2n+2}$ (n=2,3) and
(Cu,C)Ba$_{2}$Ca$_{n}$Cu$_{n+1}$O$_{3n+2}$ (n=2,3,4), where NMR
study showed that the inner CuO$_{2}$ layer(s) is less hole-doped
than the outer layers\cite{Tokunaga,Kotegawa}. We note here that
in our F0234 case, both bands are highly metallic with sharp peaks
and gap opening below T$_{c}$.

It would be very interesting to perform other experiments such as
NMR or c-axis optics to test whether this system, with the FS
areas quite far away from half-filling and with multilayer
interaction, can be better described by the band structure
language or the doped Mott insulator language described above.
Nevertheless, the isolation of the two FS' allows us to
investigate the FS dependence of the superconducting gap, which
results in important conclusions independent of the language used.

\begin{figure}
\flushleft
\includegraphics[width=0.45\textwidth]{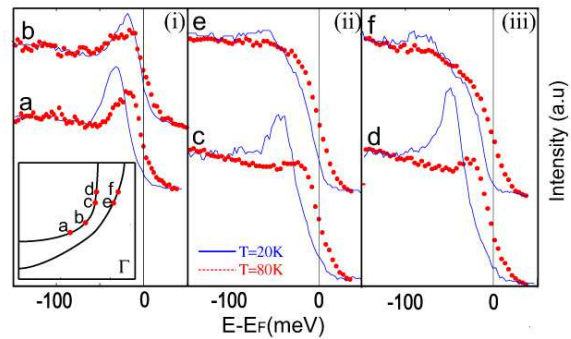}
\caption{\label{fig:epsart} Temperature dependent EDCs from points
a$\sim$f along both FS' (see inset); T=80K(red) and 20K(blue) data
are measured on two cleaves of the sample.}
\end{figure}

In Fig. 4, we present the temperature dependent EDCs from various
Fermi crossing points of both bands (see inset) that demonstrate
the opening of the superconducting gap below T$_{c}$. Due to the
simultaneous presence of band N, EDCs at points e and f appear
broader and show a hump at higher binding energy. It's clear that
the spectral intensity of both bands is pushed away from E$_{f}$
in the T=20K data compared to those measured at 80K. The strong
temperature dependence suggests the superconducting nature of the
energy gaps along both FS sheets at T$<$T$_c$. Furthermore, we
find that the gap magnitude is not only FS sheet dependent (e,f vs
c.d) but also momentum dependent within each FS sheet (a$\sim$d;
e,f)

\begin{figure}
\includegraphics[width=0.43\textwidth]{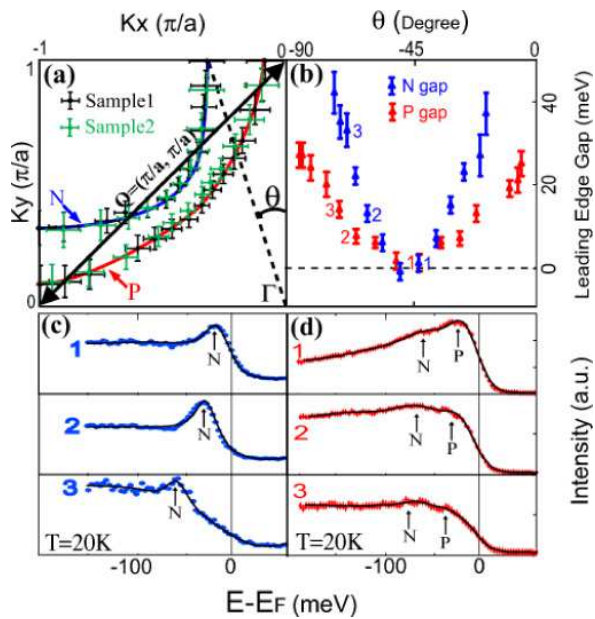}
\caption{\label{fig:epsart} (a): FS contours from two samples show
good agreement. Black double arrow shows the $(\pi/a,\pi/a)$
scattering vector. Angle $\theta$ gives the definition of
horizontal axis in (b). (b): Leading edge gap along k-space angle
from the two FS contours. (c,d): Three representative EDCs from
both FS contours (labelled as 1-3, blue for band N and red for P
respectively, measured at T=20K) that give the corresponding
points in (b). Black lines superimposed on EDCs are empirical fits
to the data; the fit peaks' positions from both bands are
indicated by black arrows.}
\end{figure}

The FS contours and superconducting gaps along them are summarized
in Fig.5. To avoid model related complexity, we present the
leading edge gap\cite{Andrea_review,ZX_gap} along both FS
contours. While the leading edge method is known to underestimate
the absolute gap magnitude, it does not affect the relative
comparison. Fig.5b gives the gap values as a function of k-space
angle as defined in Fig.5a; and Fig.5(c,d) show three sample EDCs
recorded at 20K along both FS sheets (from which the corresponding
gap values in Fig.5b are extracted). From Fig.5b, we see that the
gaps along both FS sheets have d-wave symmetry as found in other
cuprates; and although the leading edge gap magnitude along the
band P FS is comparable to that of purely hole-doped materials
such as Bi2212\cite{Andrea_review,ZX_gap}, big surprise comes from
the gap magnitude of the "electron-doped" band N FS, which is
approximately two times larger than that of its hole-doped
counterpart, and an order of magnitude larger than that of purely
electron doped materials such as
Nd$_{2-x}$Ce$_{x}$CuO$_{4}$\cite{Peter_NCCO,Sato_NCCO,Huang}.
Within the framework of doped Mott insulator, this behavior of the
self-doped system contrasts strongly to the established trend of
particle-hole asymmetry for superconductivity in purely electron-
or hole-doped systems with the n-type materials having much weaker
pairing strength\cite{Peter_NCCO,Sato_NCCO,Huang}. Unlike the
Bi2212 system with bi-layer splitting where two FS' are also
observed\cite{Pasha,Borisenko_gap}, F0234 shows non-degenerated
FS' even along the nodal direction(see Fig. 2(i$\sim$iii, v), Fig.
5a) and significant gap difference associated with different FS'.
The implications of this difference have been discussed
elsewhere\cite{Wenhui}. This makes F0234 the first self-doping
high T$_{c}$ superconductor that has a pronounced FS dependence of
the superconducting gap.

The substantially larger pairing gap in the energy dispersion
along band N FS(see Fig.2,3) strongly hints that the most
important pairing force for cooper pairs is distinct from those
theories where the van Hove singularity (VHS) near
E$_{f}$\cite{Markiewicz} or antiferromagnetic $(\pi/a,\pi/a)$
scattering\cite{Chubukov} are the most important ingredients. As
seen in Fig.5a, though band N never and band P does get close to
$(\pi/a, 0)$ in k-space where the VHS provides a large phase space
that can be connected by the $(\pi/a,\pi/a)$ scattering vector,
band N nevertheless shows much larger gap. On the other hand, the
correlation between a stronger kink and a larger gap in band N may
imply a connection. While this does not exclude the possibility
that VHS and $(\pi/a,\pi/a)$ scattering from playing some role in
the superconducting pairing\cite{Oudovenko}, our finding strongly
suggests that there exists a more important mechanism in this
material. On the other hand, our finding is consistent with the
empirical trend which indicates that it is the FS' shape of the
most bonding band that correlates with T$_{c}$\cite{Pavarini}. As
an example, HgBa$_{2}$Ca$_{2}$Cu$_{3}$O$_{8}$, the superconductor
with the highest T$_{c}$(135K) to date, is predicted\cite{Singh}
to have its most bonding band's FS similar to that of band N in
F0234.

\textbf{Acknowledgements} We thank J. Zaanen, N. Nagaosa and O.K.
Andersen for stimulating discussions. The experiments were
performed at the ALS of LBNL, which is operated by the DOE's
Office of BES, Division of Material Science, with contract
DE-AC03-76SF00098. The division also provided support for the work
at SSRL with contract DE-FG03-01ER45929-A001. The work at Stanford
was supported by NSF grant DMR- 0304981. T.P.D. acknowledges
support from NSERC and ONR grant N00014-05-1-0127.

\bigskip

\end{document}